\documentclass[aps, nofootinbib, preprintnumbers, superscriptaddress,
twocolumn]{revtex4}
\usepackage{graphicx}
\usepackage{dcolumn}
\usepackage{amssymb}
\usepackage[T1]{fontenc}
\usepackage[english]{babel}
\usepackage{graphicx}
\usepackage[latin9]{inputenc}
\usepackage{mathpazo}
\usepackage{epstopdf}
\usepackage{bm}

\newcommand{\be}{\begin{equation}}
\newcommand{\ee}{\end{equation}}
\begin{document}

\title{New, efficient, and accurate high order derivative and
  dissipation operators satisfying summation by parts, and
  applications in three-dimensional multi-block evolutions}

\author{Peter Diener}
\email{diener@cct.lsu.edu}
\affiliation{Department of Physics and Astronomy, 202 Nicholson Hall,
  Louisiana State University, Baton Rouge, LA 70803, USA}
\homepage{http://relativity.phys.lsu.edu/}
\affiliation{Center for Computation and Technology, 302 Johnston Hall,
  Louisiana State University, Baton Rouge, LA 70803, USA}
\homepage{http://www.cct.lsu.edu/}

\author{Ernst Nils Dorband}
\email{dorband@cct.lsu.edu}
\affiliation{Department of Physics and Astronomy, 202 Nicholson Hall,
  Louisiana State University, Baton Rouge, LA 70803, USA}
\homepage{http://relativity.phys.lsu.edu/}
\affiliation{Center for Computation and Technology, 302 Johnston Hall,
  Louisiana State University, Baton Rouge, LA 70803, USA}
\homepage{http://www.cct.lsu.edu/}

\author{Erik Schnetter}
\email{schnetter@cct.lsu.edu}
\affiliation{Center for Computation and Technology, 302 Johnston Hall,
  Louisiana State University, Baton Rouge, LA 70803, USA}
\homepage{http://www.cct.lsu.edu/}
\affiliation{Max-Planck-Institut für Gravitationsphysik,
  Albert-Einstein-Institut, Am Mühlenberg 1, D-14476 Golm, Germany}
\homepage{http://numrel.aei.mpg.de/}

\author{Manuel Tiglio}
\email{tiglio@cct.lsu.edu}
\affiliation{Department of Physics and Astronomy, 202 Nicholson Hall,
  Louisiana State University, Baton Rouge, LA 70803, USA}
\homepage{http://relativity.phys.lsu.edu/}
\affiliation{Center for Computation and Technology, 302 Johnston Hall,
  Louisiana State University, Baton Rouge, LA 70803, USA}
\homepage{http://www.cct.lsu.edu/}

\date{November 30, 2005}

\preprint{LSU-REL-113005}
\preprint{AEI-2005-175}

\begin{abstract}
  We construct new, efficient, and accurate high-order  finite
  differencing operators which satisfy summation by parts.  Since these
  operators are not uniquely defined, we consider several optimization
  criteria: minimizing the bandwidth, the truncation
  error on the boundary points, the spectral radius, or a combination
  of these.  We
  examine in detail a set of operators that are up to tenth order accurate in
  the interior,
  and we surprisingly find that a combination of these 
  optimizations can improve the operators' spectral radius and
  accuracy by orders of magnitude in certain cases.
  We also construct high-order dissipation
  operators that are compatible with these new finite difference operators and
  which are semi-definite with respect to the
  appropriate summation by parts scalar product.
  We test the stability
  and accuracy of these new difference and dissipation operators by evolving a
  three-dimensional scalar wave equation
  on a spherical domain consisting of seven blocks, each discretized
  with a structured grid, and connected through penalty boundary
  conditions.  
\end{abstract}

\maketitle

\section{Introduction}

Kreiss and Scherer proposed quite some time ago \cite{Kreiss1974a,Kreiss1977a} a powerful way of
constructing linearly stable schemes for solving evolution partial differential
equations which admit an energy estimate at the continuum,  through the use of
difference operators satisfying summation by parts (SBP). These operators
essentially make it possible, up to boundary terms, to derive estimates analogous to the continuum
ones. While the latter guarantee well posedness, their
discrete counterparts guarantee numerical stability. The boundary
terms left after SBP can be controlled by, for example, orthogonal
projections \cite{Olsson1995a,Olsson1995b,Olsson1995c}, penalty terms
\cite{Carpenter1994a}, or a combination of them \cite{Gustafsson98} (see
\cite{Mattsson2003a,Strand1996} for a comparison between these
methods). Furthermore, SBP difference
operators and penalty techniques have been rather recently combined to construct stable schemes
of arbitrary high order for multi-block simulations
\cite{Carpenter1999a,Nordstrom2001a}.
These are simulations where the domain is broken into different sub-domains
which are ``glued'' together through an appropriate interface treatment,
in this case penalty terms.
This semi-structured approach allows for non-trivial geometries while at the same
time ensuring stability for schemes of arbitrary high order using derivatives 
satisfying SBP.

Systems which have smooth solutions (that is, without shocks), such as the Einstein vacuum
equations (see, for
example, \cite{Reula98a}), are ideal for
using high order methods. Furthermore, in numerical relativity one typically
deals with non-trivial topologies and the need for smooth boundaries. Although
there are proofs for particular systems in non-smooth domains, 
proofs of well posedness for the initial-boundary value problem for
general symmetric hyperbolic systems usually require smooth outer boundaries
\cite{Rauch1985,Secchi1996a,Secchi1996b}.

Multi-block domains are also more efficient than single-block ones, as they
can be chosen to adapt to particular situations. For instance, 
 they can be made to mimic spherical coordinates which
automatically reduce the angular resolution at large radii (this
allowed, for example, studying late time behavior in a rotating black hole
background in full three-dimensions, placing the outer boundary at
very large distances with modest computational resources
\cite{Lehner2005a}), and one
can also reduce the radial resolution (e.g.\ logarithmically).
Last, the need for non-trivial topologies includes the particular but
very important case of black hole excision, where the black hole singularity
is removed from the computational domain.  In sum, the penalty
multi-block approach combined with high order SBP operators appears to
be promising for simulating Einstein's equations.

In principle, modulo the tedious but
straightforward symbolic manipulation algebra needed to construct high order difference operators satisfying SBP,
one can systematically generate in this way stable multi-block schemes of arbitrary high order.
However, it turns out that high order operators satisfying SBP are highly
non-unique, the higher their order the higher their non-uniqueness.
There is great variation among the properties of these operators, and
for reasons that we discuss below, much care has to be taken in
choosing the stencils if explicit time integration schemes are used.
One
approach could involve choosing operators with minimum bandwidth, as
they reduce the number of operations. Unfortunately,
in some cases the resulting operator leads to
an amplification matrix with a very large spectral radius (which has already been
pointed out in \cite{Lehner2005a} and \cite{Svard2005a}); when using explicit
schemes to integrate in time, this translates into a very small Courant limit. One can
do much better by attempting to minimize  the spectral radius of the complete
operator, rather than its bandwidth. This in some cases leads to a 
Courant limit {\em two orders of magnitude} larger as compared to the minimum
bandwidth case. One can sometimes do even
better: another feature to take into account
is the amplitude of the truncation errors at and close to boundaries.
 As we will show, in some cases one can decrease them by {\em orders of magnitude} while keeping the
spectral radius small.
What we have just briefly discussed is one of the goals of this
paper, namely to explicitly construct
efficient and accurate high order SBP difference operators, and compare
the above different criteria that can be used in their construction. We
consider both diagonal and restricted full (non-diagonal) norm based
operators; in the first (second) case up to order ten (eight) in the interior.

In many cases of interest, particularly in non-linear ones, one might
want to add a small amount of artificial dissipation to the
problem.  In order not to spoil the available energy estimates, the
dissipation operator has to be negative semi-definite with respect to the SBP
scalar product. This is not just a technical detail. As we will discuss below, in
certain cases of interest the use of simple dissipation operators that do not satisfy
this property (e.g.\ standard Kreiss--Oliger dissipation in the
interior and no dissipation near boundaries, a choice commonly used in some
applications) cannot get rid of some instabilities, while better
 dissipation operators do. Even if there are no instabilities, a
 dissipation operator that is non-zero close to boundaries is very useful if
 one wants to smooth out aspects of the solution propagating through the
 multi-block interfaces. Mattsson et al.\ \cite{Mattsson2004a} have recently presented a way of
constructing dissipation operators that are indeed negative semi-definite for
arbitrary SBP scalar products, and which extend all the way up to the
boundaries. Following this prescription one can
construct, for any difference
operator of arbitrary high order satisfying SBP, an associated dissipation
{\em up to the very boundary points} in a systematic way. In
this paper we do so explicitly, for each of the efficient and accurate high
order derivatives that we present. This is the second goal of our paper.

The third and final goal is to test these new derivative and
dissipation operators in three-dimensional multi-block
simulations, making use of the penalty method to handle
interfaces, as described in \cite{Carpenter1999a,Nordstrom2001a}. Because of the
challenge involved in
achieving stability for very high order schemes in the presence of interfaces, multi-block
domains present an ideal setting for testing the new derivative and
dissipation operators that we here construct. While black hole excision is one of our
main motivations for using multiple blocks, we will report here on
simulations on a domain that is useful for scenarios that do not
involve black hole excision, but still need a smooth (e.g.\ spherical) outer boundary. This grid
structure should be useful for studies of wave phenomena at large distances
from the source, gravitational collapse, or
Friedrich's conformal approach, where the spacetime is
compactified, and null infinity is brought to a finite computational
distance (see e.g.\ \cite{Friedrich:2002xz}).
In order to isolate testing numerical stability, accuracy, and
efficiency of the new high order derivative and dissipation operators 
 from gauge problems and continuum instabilities
typically found in many formulations of
the Einstein equations, we perform these
tests in this paper using a simpler three-dimensional system ---a
massless scalar field---
and we will report on evolutions of the Einstein equations elsewhere.

This paper is organized as follows.  
In Section \ref{setup} we introduce our notation, review shortly the
penalty method, discuss the relative merits of SBP finite differencing
operators based on diagonal and on non-diagonal norms, and summarize
the construction of dissipation operators of Mattsson et al.\
\cite{Mattsson2004a}.
In Section \ref{sbp}, we explain the different strategies that we use
in constructing the new derivative operators, namely their bandwidth,
their spectral radius, and their truncation error.
We introduce our example system of evolution equations in Section
\ref{eqs}, where we also describe the type of three-dimensional
multi-block domain that we use to test the new difference and
dissipation operators.
We describe our new operators corresponding to diagonal and restricted
full norms in Section \ref{ops}, discussing their properties and
comparing their accuracies in numerical tests.
Finally, Section \ref{end} closes with some remarks about possible
future research directions.

\section{SBP derivative and dissipation operators with a high order of
  accuracy} \label{setup}

\subsection{SBP and penalties}\label{normdefs}
In this subsection we briefly summarize Section 2 of  \cite{Lehner2005a}.
We do so essentially to fix our notation, for more details see that reference.
Consider a computational domain $[a,b]$ and a discrete grid consisting
of points $i=1\ldots n$ and grid spacing $h$.
A difference
operator $D$ is said to satisfy SBP on that domain with respect to a positive
definite scalar product $\Sigma$ (defined by its coefficients
$\sigma_{ij}$)
\begin{equation}
\langle u,v\rangle = h\sum_{i,j=1}^n u_i v_j \sigma_{ij} \label{prod} \, ,
\end{equation}
if the property
\begin{equation}
\langle u,Dv \rangle + \langle v,Du \rangle = \left(uv\right)|_a^b
\end{equation}
holds for all grid functions $u$ and $v$. Similar definitions can be introduced for
two (and higher) dimensional domains. The scalar product or norm is said to
be \emph{diagonal} if
\begin{equation}
\sigma_{ij} = \sigma_{ii} \delta_{ij} \;\textrm{,}
\end{equation}
that is, if $\sigma_{ij}$ is diagonal.  It is called \emph{restricted
  full} if
\begin{equation}
\sigma_{i_bj} = \sigma_{i_bi_b} \delta_{i_bj} \;\textrm{,}
\end{equation}
that is, if $\sigma_{ij}$ is diagonal on the boundary, but may be
non-diagonal (full) in the interior.  $i_b \in \{1,n\}$ denote boundary point
indices.

We now briefly highlight through a simple example the main features of the penalty method for multi-block evolutions, for
more details see \cite{Carpenter1999a,Nordstrom2001a}. We assume that the norm is either diagonal or
restricted full, since these are the cases we
actually consider later in this paper.

The simple example we wish to consider is the advection equation,
$$
\partial_t u = \Lambda \partial_x u
$$
in the spatial interval $(-\infty,+\infty)$ with appropriate fall-off
conditions at
infinity, 
and two grids: a \emph{left grid} covering $(-\infty,0]$, and
a \emph{right grid} covering $[0,+\infty)$. We refer to the grid function $u$ on each grid by
$u^l$ and $u^r$, corresponding to the left and right grids, respectively.
The problem is discretized using
grid spacings $h^l,h^r$ on the left and right grids ---not necessarily equal--- and
difference operators $D^l,D^r$ satisfying SBP with respect to scalar products
given by the weights $\sigma^l, \sigma^r$ \emph{on their individual grids}.
That is, these scalar products are defined through
\begin{equation}
\langle u^l,v^l \rangle =  h^l\sum_{i,j=-\infty}^{0} \sigma^l_{ij} u^l_i v^l_j  \;\;\; , \;\;\;
\langle u^r,v^r \rangle =  h^r\sum_{i,j=0}^{+\infty} \sigma^r_{ij}  u^r_i v^r_j
\end{equation}

The semi-discrete equations are written as
\begin{eqnarray}
\partial_t u^l_i &=& \Lambda D^l u^l_i + \frac{\delta_{i,0} S^l}{h^l\sigma^l_{00}}(u^r_0-u^l_0) \label{advl} \;\textrm{,}\\
\partial_t u^r_i &=& \Lambda D^r u^r_i + \frac{\delta_{i,0} S^r}{h^r\sigma^r_{00}}(u^l_0-u^r_0) \label{advr} \;\textrm{.}
\end{eqnarray}
In the fully non-diagonal case the treatment is slightly more
complicated, therefore we consider here only the diagonal and the
restricted full cases.

One can derive an energy estimate and therefore guarantee stability if two
conditions are satisfied. One of them is 
$\Lambda + S_r - S_l = 0$. The other one imposes an additional
constraint on the values of $S_l$
and $S_r$:
\begin{itemize}
\item \emph{Positive $\Lambda$:}
\be
S_l = \Lambda + \delta, \;\;\; S_r = \delta, \;\;\; \mbox{ with } \delta \geq
- \frac{\Lambda}{2} \label{pos_lam}
\ee
\item \emph{Negative $\Lambda$:}
\be
S_r = -\Lambda + \delta, \;\;\; S_l = \delta, \;\;\; \mbox{ with } \delta \geq
 \frac{\Lambda}{2} \label{neg_lam}
\ee
\item \emph{Vanishing $\Lambda$:} this can be seen as the limiting case of any of the above two.
\end{itemize}
For the minimum values of $\delta $ allowed by the above inequalities the energy estimate is the
same as for a single grid (that is, as if the interface did
not exist), while for larger values
of $\delta$ there is
damping in the energy which is proportional to the mismatch at the interface.

\subsection{Diagonal versus non-diagonal norms} \label{diag_vs_nondiag}

There are several advantages in using one-dimensional (1D) difference operators
satisfying SBP with respect to diagonal norms. One of them is
related to the fact that SBP is guaranteed to hold in several
dimensions by simply applying the 1D operator along each direction
\cite{Olsson1995a,Olsson1995b,Olsson1995c}. Another advantage is related to the following: in
order to ensure stability through an energy estimate, in many cases
one has to be able to bound the norm of the commutator between the
difference operator and the principal part of the equations \emph{for
  all resolutions}, and this is guaranteed to hold in the
diagonal case. Finally, the expressions of the operators are also
somewhat simpler when compared to non-diagonal ones. The
disadvantage, on the other hand, is that the order of accuracy at and
close to boundaries is half of that in the interior, while in the
restricted full case the operators lose only one order near boundaries
\cite{Kreiss1974a,Kreiss1977a,Strand1994a}.

Though more efficient, schemes based on non-diagonal norms
might have stability problems in the absence of dissipation. First, it is not
guaranteed that SBP holds in several dimensions if a difference
operator satisfying SBP in 1D is applied along each direction.
Second, there can be problems even in one dimension when the system
has variable coefficients, since the boundedness of the commutator
discussed above is not guaranteed to hold either. This is not a
feature inherent to finite difference (FD)-based schemes: the
boundedness of such commutator is, for example, in general not guaranteed
either for pseudo-spectral methods in the absence of filtering, even in
periodic domains, when the system has variable coefficients
\cite{Tadmor94}.  Therefore, both in the
case of pseudo-spectral methods and non-diagonal norm based FD
schemes, one is in general unable to guarantee stability in more than
one dimension, or even in 1D in the variable coefficient case, without
filtering or dissipation (see also \cite{Svard2005b}). In the problem at hand,
the question then is
whether one can stabilize the scheme through artificial dissipation,
without introducing an excessive amount of it. Below we will address this
question in detail, as well as compare diagonal based operators
to their non-diagonal counterparts.

In the diagonal case we will consider difference operators of order
two, four, six, eight, and ten in the interior (and therefore order
one, two, three, four and five, respectively, at and close to
boundaries) and denote them by
$D_{2-1},D_{4-2},D_{6-3},D_{8-4},D_{10-5}$. In the non-diagonal (restricted full)
 case,  we will consider operators of order four, six and eight in
the interior (and therefore order three, five and seven, respectively, at
and close to boundaries), and denote them by $D_{4-3},D_{6-5},D_{8-7}$. 
These operators in general are not unique. For example, in the
second order in the interior case there is a unique operator
satisfying SBP, and its norm is diagonal, the operator being what we
called $D_{2-1}$. With respect to higher order operators, the
following holds for the diagonal norm based ones: $D_{4-2}$ is
unique, while $D_{6-3}$, $D_{8-4}$ and $D_{10-5}$ comprise a one-,
three-, and ten-parameter family, respectively. In the restricted full
case, $D_{4-3}$, $D_{6-5}$ and $D_{8-7}$ have three, four and five free
parameters, respectively.

\subsection{Dissipation operators}

As pointed out in~\cite{Mattsson2004a}, adding artificial dissipation
may lead to an unstable scheme unless the dissipation operator is compatible
with the SBP derivative operator. In that reference,
the authors present a prescription for constructing
such operators, which we follow here. In short, a compatible dissipation
operator, of order $2p$ in the interior, is constructed as
\begin{equation}
A_{2p} = - \alpha\, h^{2p}\; \Sigma^{-1} D_p^T B_p D_p, \label{dismat}
\end{equation}
where $\alpha$ is a positive constant, $\Sigma$ is the
scalar product used in the construction of the SBP operator, and $D_p$
is a consistent approximation of $d^p/dx^p$ with minimal
width\footnote{``Minimal width'' means that the stencil must contain
  as few points as possible.}.  $B_p$ is the so-called \emph{boundary
  operator}.  The boundary operator is positive semi-definite and its role
is to allow boundary points to be treated differently from interior
points.  $B_p$ cannot be chosen freely, but has to follow certain
restrictions which we explain below.

For the diagonal norm operators, choosing $B_p$ to be the unit matrix
is sufficient to obtain the required $p$th order accuracy near the
boundary, which is the same accuracy as the derivative operator.

In the case of restricted full norm operators, the accuracy
requirement near the boundary is stricter.  The dissipation operator
should have order $2p-1$ at the boundary and order $2p$ in the
interior, which requires a different choice of $B_p$.  We again
follow~\cite{Mattsson2004a} and choose $B_p$ to be a diagonal matrix,
where the diagonal is the restriction onto the grid of a piecewise
smooth function. The numerical domain is divided into three regions in each
dimension; an interior part and on either side two transition regions 
containing the boundaries. The transition region has a fixed size that is
independent of the resolution. Within the transition region the function,
$B_p$, increases from $O(h^{p-1})$ at the outer boundary to a constant value
$1$ at the boundary with the interior region in such a way that the
derivatives of $B_p$ up to order $p-2$ vanish at either ends. In the interior
region the function has the constant value $1$.

For the $D_{4-3}$ operator, $B_p$ has the value
$h$ at the boundary and increases linearly to $1$ in the transition
region. For the $D_{6-5}$ operator, we use a cubic polynomial with
vanishing derivatives at either end of the transition region to
increase the value of $B_p$ from $h^2$ at the boundary to $1$ in the
interior. For the $D_{8-7}$ operator, the boundary values for the
transition region are $h^3$ and $1$, and we use a fifth order polynomial to
make the first and second derivatives vanish at either end of the transition
region.

For the constant $\alpha$ we make the choice $\alpha=2^{-2p}$, since then the
parameter used to specify the strength of the dissipation has approximately
the same allowed numerical range, independently of the order of the operator.

Note that in the diagonal case up to order eight in the interior, the scalar
product $\Sigma$ is independent
of the free parameters, so for a given order the same dissipation operator
is used for all the different operators we construct below, while for
the higher order diagonal operators and the restricted full norm operators
a unique dissipation operator has to
be constructed for each parameter choice.

\section{Optimization criteria} \label{sbp}

We start by fixing some notation.
If the accuracy of the difference operator $D$ in the interior 
is $2p$, then there are $b$ points at and
near the boundaries where the order of $D$ is only $q$.  In the
diagonal case one has $q=p$, and in the restricted full case it is
$q=2p-1$.  We call $b$ the
boundary width. The difference operator at these $b$ points
uses (up to) $s$ points to compute the derivative.  We call $s$
the boundary stencil size.

When expanding $D$ in a Taylor series one has
\begin{equation}
\left. Du \right|_{x_i} = \left. \frac{du}{dx} \right|_{x_i} + c_ih^q
\left. \frac{d^{q+1}u}{dx^{q+1}} \right|_{x_i}
\quad\mathrm{for}\;i=1\ldots b
\end{equation}
where $h$ is the grid spacing and $x_i=ih$.  We call
$c_i$ the \emph{error coefficients}.

In what follows, we consider
three cases for each family of operators of a given order, denoted by:
\begin{itemize}
\item{\it Minimum bandwidth:} If there are $n$ free parameters, it is always
  possible to set $n$ of the derivative coefficients to zero, thereby
  minimizing the computational cost of evaluating the derivatives in the
  boundary region. 
\item{\it Minimum spectral radius:} In this case, we calculate
  numerically the eigenvalues of the amplification matrix for a test
  problem, and choose the parameters to minimize the largest
  eigenvalue.
\item{\it Minimum ABTE:} We minimize the average boundary truncation
  error (ABTE), which we define as
\begin{equation}
\textrm{ABTE}:=\left( \frac{1}{b}\sum_{i=1}^bc_i^2  \right)^{1/2}. \label{TE}
\end{equation}
\end{itemize}

The test problem that we use to compute the spectral radius of the
amplification matrix is the same
one that was used in \cite{Lehner2005a}: an
advection equation propagating in a periodic domain. Periodicity is
enforced through an artificial interface boundary via penalties.%
\footnote{Truly periodic domains (that is, without an interface) do not
  require boundary derivative operators, and therefore do not constitute a
  useful test here.} We use a penalty parameter $\delta = -1/2$ (see Section
\ref{setup}), which means
that the semi-discrete energy is strictly preserved, and that the amplification
matrix is anti-Hermitian, and therefore the real part of all
eigenvalues is zero.%
\footnote{As a side remark: we actually compute the eigenvalues of the
  amplification
  matrix multiplied by the grid spacing $h$.}

Another option
\cite{Svard2005a} would be to compute the spectral radius of the discrete
difference operator itself. In this case, the spectrum is in general not purely
imaginary, since the boundary conditions have not been imposed yet. In
practice we have found, though, that both approaches lead to similar
operators, in the sense that
a derivative operator with small spectral radius usually also leads to
amplification
matrices for the above test problem with small spectral radii as well. 

It is worth pointing out that for the diagonal operator case the bandwidth
and the ABTE can be globally minimized by analytically choosing
the parameters, since the ABTE is a quadratic function of the
parameters and therefore has a global minimum.
This is not the case for the spectral radius.
Therefore, when we refer to minimizing the spectral radius, we perform a
numerical minimization and do not claim that we have actually found a global
minimum.

\section{Example evolution system and multi-block domain setup}
\label{eqs}

We test each of the new derivative operators and their associated
dissipation operators through 3D multi-block simulations. 
In these simulations we solve the scalar wave equation
\begin{equation}
\partial_{tt} \phi = \Delta \phi
\end{equation}
in static, curvilinear coordinates. This equation can be
reformulated as
\begin{eqnarray}
\label{eq:waveequation}
\partial_t \phi &=& \Pi \label{phi_dot}\\
\partial_t \Pi &=& \gamma^{-1/2} D_i \left ( \gamma^{1/2} \gamma^{ij}
  d_j \right) \label{pi_dot}\\
\partial_t d_i &=& D_i \Pi \;,  \label{d_dot}
\end{eqnarray}
where $\gamma_{ij}$ is the Euclidean metric in curvilinear coordinates,
$\gamma=\det \gamma_{ij}$ its determinant, and $\gamma^{ij}$ its
inverse.  Here we use the Einstein summation convention,
implicitly summing over repeated indices.

The advantage of this particular form of the equations is that it
defines a strictly stable discretization in the diagonal case,
in the sense that one can
show by just using SBP (i.e., without needing a discrete Leibniz rule,
which in general does
not hold) that there is a semi-discrete energy $E$ which is preserved in time
for any difference operator $D$ satisfying SBP.%
\footnote{That is, the energy is preserved modulo boundary conditions,
  which can inject or remove energy from the system.  For example, if
  maximally dissipative boundary conditions are used, the energy
  actually decreases as a function of time.}\@
This energy is given by
\begin{equation}
  E = \frac{1}{2} \int \left( \Pi^2 + \gamma^{ij} d_i d_j \right) \,
  \gamma^{1/2} \; dV .
\end{equation}

\begin{figure}
\begin{center}
\includegraphics[width=0.45\textwidth]{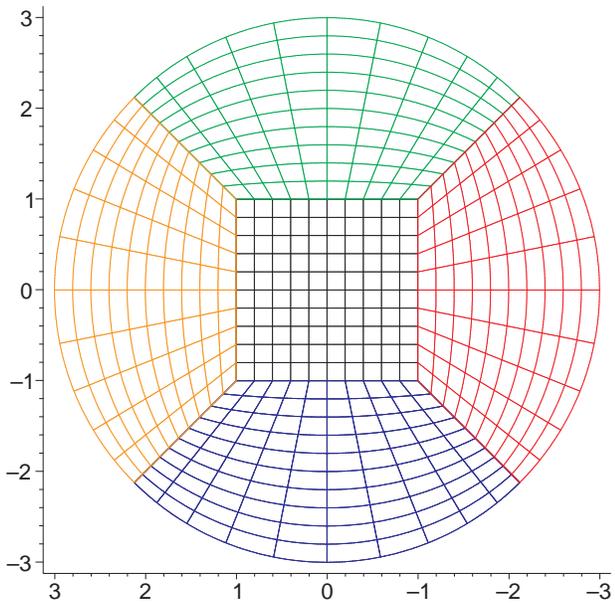}
\caption{An equatorial cut of the 3D multi-block structure used in
  the simulations of this paper.}
\label{grid}
\end{center}
\end{figure}

The geometry of the computational domain in the three-dimensional (3D)
simulations
that we show below is the interior of a sphere.
In order to avoid the singularities
at the origin and poles that spherical coordinates have, we cover
the domain with seven blocks: surrounding the origin we use a
Cartesian, cubic block, which is matched (at each face) to a set of
six blocks which are a deformation of the cubed-sphere coordinates
used in \cite{Lehner2005a}. Figure \ref{grid} shows an
equatorial cut of our 3D grid structure, for $11^3$ points on each
block.  The blocks touch and have one grid point in common at the
faces.  The grid lines are continuous across interfaces, but in general
not smooth.

Each block uses coordinates $a,b,c$. For the inner block these are the
standard Cartesian ones: $a=x,b=y,c=z$, while for the six outer
blocks they are defined as follows. The ``radial'' coordinate
$c\in[-1,1]$ is defined by inverting the relationship \be r =
\frac{1}{2} \left[r_0(1-c) + r_1(1+c)  \right] \label{rc} \ee (where
$r=\sqrt{x^2+y^2+z^2}$). The ``angular'' coordinates,
$a,b\in[-1,1]$, are in turn defined through
\begin{itemize}
\item Neighborhood of positive $x$ axis:\\ $x=r/F,\quad y=rb/F,\quad z=ar/F$
\item Neighborhood of positive $y$ axis:\\ $x=-br/F,\quad y=r/F,\quad z=ar/F$
\item Neighborhood of negative $x$ axis:\\ $x=-r/F,\quad y=-rb/F,\quad z=ar/F$
\item Neighborhood of negative $y$ axis:\\ $x=br/F,\quad y=-r/F,\quad z=ar/F$
\item Neighborhood of positive $z$ axis:\\ $x=-ar/F,\quad y= rb/F,\quad z=r/F$
\item Neighborhood of negative $z$ axis:\\ $x=ar/F,\quad y= rb/F,\quad z=-r/F$
\end{itemize}
where $r$ is written in terms of $c$ through (\ref{rc}), and
\begin{eqnarray}
F & := & \left(\frac{(r_1-r) + (r-r_0)E}{r_1-r_0}\right)^{1/2}
\end{eqnarray}
with $E=1+a^2+b^2$. The surface $c=1$ corresponds to the spherical
outer boundary (of radius $r=r_1$), while $c=-1$ corresponds to the
cubic interface boundary matched to a cube of length $2r_0$ on each
direction. The grid structure of Figure \ref{grid}
corresponds to $r_0=1$, $r_1=3$.

To get an accurate measure of numerical errors, we evolve initial data for
which there exists a simple analytic solution of (\ref{eq:waveequation}) for all
times, against which we can compare the numerical results. We choose
initial data
\begin{eqnarray}
\phi(t=0) &=& A \cos(2 \pi  \bm{k} \cdot \bm{x}) \\
\Pi(t=0) &=& - 2 \pi A |\bm{k}| \sin(2 \pi \bm{k} \cdot \bm{x}) \\
d_i(t=0) &=& - 2 \pi A k_i \sin(2 \pi \bm{k} \cdot \bm{x}).
\end{eqnarray} 
The analytic solution for this setup is a plane wave with constant amplitude
$A$ traveling  through the grid in the direction of the
vector $\bm{k}$.
In all the simulations that we present below we use $A=1.0$ and
$\bm{k}=(0.2,0.2,0.2)$, 
i.e., the wave is traveling in the direction of the main diagonal.
Consistent and stable outer boundary conditions are imposed through penalty
terms by penalizing the incoming characteristic modes with the
difference to the exact
solution, as introduced in \cite{Carpenter1994a}.
As mentioned above, we set $r_0=1$ and $r_1=3$ in our block system,
placing the outer boundary at $R=3$.

We use resolutions from $h = \Delta a = \Delta b = \Delta c = 0.1$
($21^3$ grid points per block) up to $h = 0.0125$ ($161^3$ grid points
per block).  Our highest resolution corresponds to about $400$ grid
points per wave length.  (This figure depends on which part of which
block one looks at, since the wave does not propagate everywhere along
grid lines.)
We use that many grid points per wave length ---or, equivalently put,
we use such a large wave length--- because we are interested in high
accuracy.  Decreasing the wave length is resolution-wise equivalent to
using fewer grid points per block, which is a case we study with our
coarse resolution.
It would be interesting to study the effect of very short length
features onto the stability of the system, i.e., to use a wave length
that cannot be well resolved any more.  Discrete stability guarantees
in this case that the evolution remains stable, and we assume that a
suitable amount of artificial dissipation can help in the non-linear
case when there is no known energy estimate.
The size of the time step is chosen
to be proportional to the minimal grid spacing in local coordinates
$\Delta t = \lambda\, \min (\Delta a, \Delta b, \Delta c)$ with the
Courant factor $\lambda$.   Unless otherwise
stated, we use $\lambda = 0.25$.
For the penalty terms, we used $\delta=0$ everywhere.
For all the runs with dissipation the strength was chosen to be
$\epsilon = 0.4$ (see Appendix \ref{dis_ap}).
For the dissipation operators based on a non-diagonal norm, we choose
the transition region to be $30$\% of the domain size.
Note that except for the $D_{6-5}$ operator (where dissipation is essential
in order to stabilize the operator) the differences between results with
and without dissipation are so small, that we only show the results obtained
without dissipation.
In all cases we calculate the
error in the numerical solution for $\phi(\bm{x},t)$ with respect to
the exact analytical solution.

We have implemented our code in the Cactus framework
\cite{Goodale02a,cactusweb1} using the Carpet infrastructure
\cite{Schnetter-etal-03b,carpetweb}.

\section{Operators}\label{ops}
\subsection{Operators based on diagonal norms}\label{opdiag}
\subsubsection{Operator properties}

We consider first operators that are based on a diagonal norm, since
this is the easier case.
We examine here the operators $D_{6-3}$, $D_{8-4}$, and $D_{10-5}$.
These operators have 6, 8,
and 11 boundary points, respectively, and their maximum stencil
sizes are 9, 12, and 16 points, respectively.\footnote{We
expected the $D_{10-5}$
operator to have 10 boundary points and a maximum stencil size
of 15 points, but this did not result in a positive definite norm.} The
operators $D_{2-1}$ and $D_{4-2}$ are also based on a diagonal norm.
They are unique and have been examined in \cite{Lehner2005a}. For completeness
we list their properties here as well.

The $D_{2-1}$ operator formally has two boundary points and a maximal stencil
size of three points. However, the stencil for the second boundary point is
the same as the interior centered stencil, so in practice it has only one
boundary point with a stencil size of two points. The spectral radius
and error coefficient are listed in table \ref{tab:2-1}.
\begin{table}
\caption{Properties of the $D_{2-1}$ operator.}
\begin{ruledtabular}
\begin{tabular}{l|d}
Operator & \multicolumn{1}{c}{Unique} \\\hline
Spectral radius & 1.414 \\
ABTE & 0.25 \\
$c_1$ & 0.5
\end{tabular}
\label{tab:2-1}
\end{ruledtabular}
\end{table}
The $D_{4-2}$ operator has four boundary points and a maximal stencil size of
six points.  We list its properties in table \ref{tab:4-2}.
\begin{table}
\caption{Properties of the $D_{4-2}$ operator.}
\begin{ruledtabular}
\begin{tabular}{l|d}
Operator & \multicolumn{1}{c}{Unique} \\\hline
Spectral radius & 1.936 \\
ABTE & 0.2276 \\
$c_1$ & -0.4215  \\
$c_2$ & 0.1666  \\
$c_3$ & -0.0193 \\
$c_4$ & -0.037
\end{tabular}
\label{tab:4-2}
\end{ruledtabular}
\end{table}

The family of $D_{6-3}$ operators has one free parameter.  The
resulting norm is positive definite, and is independent of this
parameter, hence the parameter can be freely chosen.  For this
operator there are very small numerical differences in the spectral
radius and in the truncation error coefficients between the three
different cases where the bandwidth, the spectral radius and the
 average boundary truncation error are respectively minimized, as can be seen
in table \ref{tab:6-3-comparison}.
\begin{table}
\caption{Properties of the diagonal norm $D_{6-3}$ operators.}
\begin{ruledtabular}
\begin{tabular}{l|ddd}
& \multicolumn{1}{c}{Minimum} & \multicolumn{1}{c}{Minimum} & \multicolumn{1}{c}{Minimum} \\
Operator & \multicolumn{1}{c}{bandwidth} & \multicolumn{1}{c}{spectral radius} & \multicolumn{1}{c}{ABTE} \\\hline
Spectral radius & 2.1287 & 2.1077 & 2.1082 \\
ABTE &  0.2716 & 0.2563 &  0.2558  \\
$c_1$ & 0.5008 & 0.5436 & 0.5374 \\
$c_2$ & -0.1854 & -0.2340 & -0.2270 \\
$c_3$ & -0.2144 & 0.0012 & -0.0300 \\
$c_4$ & 0.3067 & 0.1977 & 0.2135 \\
$c_5$ & -0.1288 & -0.0546 & -0.0654 \\
$c_6$ & -0.0286 & -0.0419 & -0.0400
\end{tabular}
\label{tab:6-3-comparison}
\end{ruledtabular}
\end{table}

Based on those small differences one would expect that there should
not be much of a difference in terms of accuracy among these three
different cases in practical simulations.  However, it turns out that
the minimum bandwidth operator in practice
leads to very different solution errors
compared to the minimum ABTE operator, and that the latter
is to be preferred.  We did not implement the minimum spectral radius
operator, since the difference in spectral radius is minimal.

The family of $D_{8-4}$ operators has three free parameters; as in the
previous case, the norm is positive definite and independent of the
parameters, which can therefore be freely chosen.  We again
investigate the properties of the operators obtained by minimizing the
bandwidth, the spectral radius, and the ABTE.
Interestingly,
we find out that there is a one parameter family of operators that
minimizes the ABTE.  Therefore, in the minimum ABTE case we make use of
this freedom and also decrease the spectral radius as much as
possible.  The results are shown in table \ref{tab:8-4-comparison}.
\begin{table}
\caption{Properties of the diagonal norm $D_{8-4}$ operators.}
\begin{ruledtabular}
\begin{tabular}{l|ddd}
& \multicolumn{1}{c}{Minimum} & \multicolumn{1}{c}{Minimum} & \multicolumn{1}{c}{Minimum} \\
Operator & \multicolumn{1}{c}{bandwidth} & \multicolumn{1}{c}{spectral radius} & \multicolumn{1}{c}{ABTE} \\\hline
Spectral radius & 16.0376 & 2.229 & 2.231 \\
ABTE & 1.2241 &   0.3993  & 0.3474 \\
$c_1$ & -0.5878 & -0.8277 & -0.8086 \\
$c_2$ & 0.1068 & 0.3682 & 0.3439 \\
$c_3$ & 3.1427 & -0.3819 & 0.0228 \\
$c_4$ & -0.7918 & -0.2186 & -0.3086 \\
$c_5$ & 0.9886 & -0.3412 & 0.0225 \\
$c_6$ & 0.3304 & 0.3619 & 0.2970 \\
$c_7$ & -0.1995 & -0.1097 & -0.0823 \\
$c_8$ & -0.0211 & -0.0465 & -0.0497
\end{tabular}
\label{tab:8-4-comparison}
\end{ruledtabular}
\end{table}

In this case, the minimum bandwidth operator is quite unacceptable due
to its large spectral radius.  (For this reason we did not even
implement it.)  The error coefficients are also quite large compared
to the two other cases.  The differences in error coefficients between
the minimum spectral radius operator and the minimum ABTE operator
might appear quite small, but as demonstrated in figure \ref{84comp}
below, there is almost of factor of two in the magnitude of the error
in our numerical tests (see below).

The family of $D_{10-5}$ operators turns out to be different from the
lower order cases. The conditions the SBP property impose on the norm do not 
yield a positive definite solution with a boundary width of 10 points. When
using a boundary width of 11 points instead, there is a free parameter in
the norm, which for a very narrow range of values does allow
it to be positive definite.
We choose this parameter to be approximately in the center of
the allowed range. With the larger boundary width, there are 10 free
parameters in the difference operator. Fixing these to give a minimal bandwidth operator results in
an operator with a large ABTE (20.534) and very large spectral radius (995.9)
that is not of practical use. Minimizing the spectral radius in the full
10 dimensional parameter space turned out to be very difficult because
the largest imaginary eigenvalue does not vary smoothly with the parameters.
Instead we attempted to minimize the average magnitude of all the eigenvalues,
however the resulting operator turned out to have a slightly larger spectral
radius than the minimum ABTE operator considered next and were
therefore not implemented and tested.
Minimizing the ABTE instead fixes four of the
ten parameters and the remaining six can then be used to minimize the spectral
radius. This results in an operator with a ABTE of 0.7661 and spectral radius of
2.240. When used in practice, it turns out that this operator at moderate
resolutions has rather large errors compared to the corresponding $D_{8-4}$ case. The
errors can be reduced by a factor of about 3 by adding artificial dissipation,
indicating that the errors, though not growing in time, are dominated by high frequency noise from the
boundary derivative operators.
Only at the highest resolution considered in the tests of
this paper is there an advantage in using the
$D_{10-5}$ operators. We therefore do not pursue them further in this paper,
though we might consider their use in other applications if we need
higher resolutions.

For completeness we list the properties of the minimum bandwidth and ABTE
$D_{10-5}$ operators in table~\ref{tab:10-5-comparison}.
\begin{table}
\caption{Properties of the diagonal norm $D_{10-5}$ operators.}
\begin{ruledtabular}
\begin{tabular}{l|dd}
& \multicolumn{1}{c}{Minimum} & \multicolumn{1}{c}{Minimum} \\
Operator & \multicolumn{1}{c}{bandwidth} & \multicolumn{1}{c}{ABTE} \\\hline
Spectral radius & 995.9 & 2.240 \\
ABTE & 20.534 & 0.7661 \\
$c_1$ & 0.7270 & 2.0379 \\
$c_2$ & 0.3034 & -0.8545 \\
$c_3$ & -10.5442 & -0.0898 \\
$c_4$ & 0.2690 & 0.7250 \\
$c_5$ & 4.8373 & 0.2889 \\
$c_6$ & -62.3907 & -0.0154 \\
$c_7$ & -1.3649 & -0.6429 \\
$c_8$ & 24.6628 & 0.0293 \\
$c_9$ & 0.1045 & 0.7073\\
$c_{10}$ & -0.4197 & -0.1411 \\
$c_{11}$ & -0.0243 & -0.1469
\end{tabular}
\label{tab:10-5-comparison}
\end{ruledtabular}
\end{table}

\subsubsection{Numerical tests}\label{numdiag}

In the diagonal case we do not need to add dissipation to the
equations which we solve here, since our semi-discrete discretization
is strictly stable, as discussed in Section \ref{eqs}. 
This means
that the errors cannot grow as a function of time at a fixed
resolution (i.e., at the semi-discrete level).  However, following
\cite{Mattsson2004a} we have constructed corresponding dissipation
operators for these derivatives for future use in non-linear problems
\cite{Zink2005b}.

\begin{figure}
\begin{center}
\includegraphics[width=0.45\textwidth]{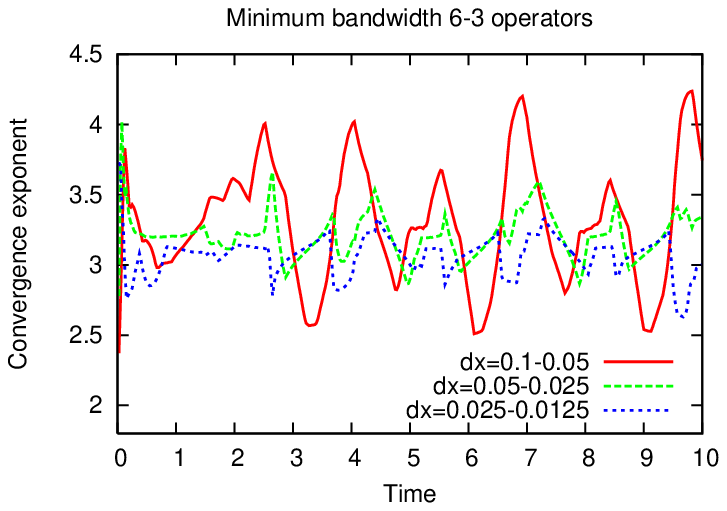}
\includegraphics[width=0.45\textwidth]{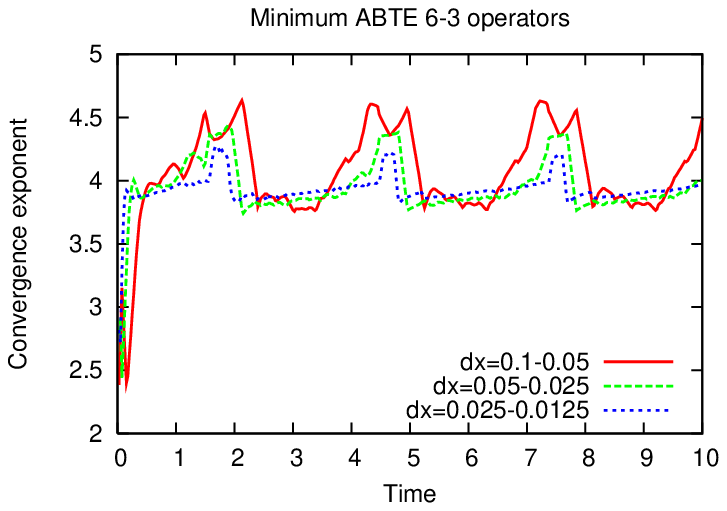}
\caption{Convergence exponents for the minimum bandwidth (top) and the
  minimum ABTE (bottom) $D_{6-3}$ operators.}
\label{63conv}
\end{center}
\end{figure}

\begin{figure}
\begin{center}
\includegraphics[width=0.45\textwidth]{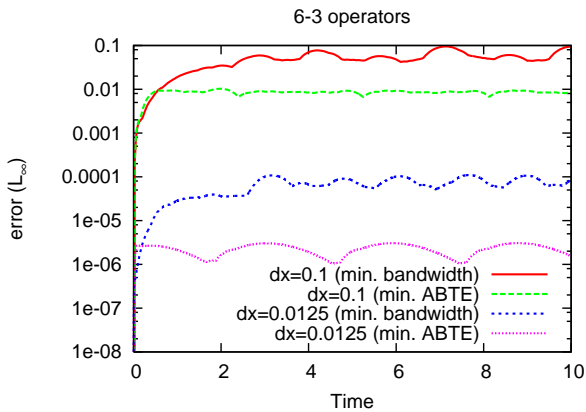}
\caption{Accuracy comparison between the $D_{6-3}$ operators of the previous figure. 
For the highest resolution that we used, the
errors with the minimum average boundary truncation error operator are around {\em two orders of magnitude
  smaller} than those obtained with the minimum bandwidth one.}
\label{63comp}
\end{center}
\end{figure}

\paragraph{The operator $D_{6-3}$.}\label{D_6_3}
Figure \ref{63conv} shows the results of convergence tests in the $L_{\infty}$
norm for the $D_{6-3}$ case, for both the
minimum bandwidth and the minimum ABTE operators. 
We define the convergence exponent $m$ as
\begin{eqnarray}
  m & = & \frac{\log \frac{E_1}{E_2}}{\log \frac{h_1}{h_2}}
\end{eqnarray}
where $E_1$ and $E_2$ are solution errors and $h_1$ and $h_2$ the
corresponding resolutions.
In the minimum bandwidth case the convergence exponent gets close to
three as resolution is increased, 
i.e., the order is being dominated by boundary (outer, interface, or both)
effects.
On the other hand, one can see from the figure 
that in the minimum ABTE case we do get a global convergence exponent that
gets quite close to four when resolution is increased.
Figure \ref{63comp} shows an accuracy comparison between these two
operators, for the coarsest and highest resolutions used in the previous
plots, displaying the errors with respect
to the exact solution in the $L_{\infty}$ norm. The improvement is quite
impressive: for the highest resolution that we used the error 
with the minimum ABTE operator is around {\em two orders of magnitude smaller}.


We conjecture that this is caused by larger truncation errors for this
operator near boundaries.  This is unfortunately not immediately
evident when looking at the boundary error coefficients $c_i$ or the
ABTE.  However, three of the inner four error coefficients have absolute
values less than $0.1$ for the optimized operator, whereas this is the
case for only 1 error coefficient for the standard operator.  The fact
that the accuracy is also higher with the optimized operator also
points to the fact that the standard operator introduces somehow a
much larger error.

As a summary, the minimum ABTE operator is the preferred choice in this case,
the minimum bandwidth $D_{6-3}$ operator does not have a large spectral radius in
comparison, but it does have much larger truncation error coefficients. 

\begin{figure}
\begin{center}
\includegraphics[width=0.45\textwidth]{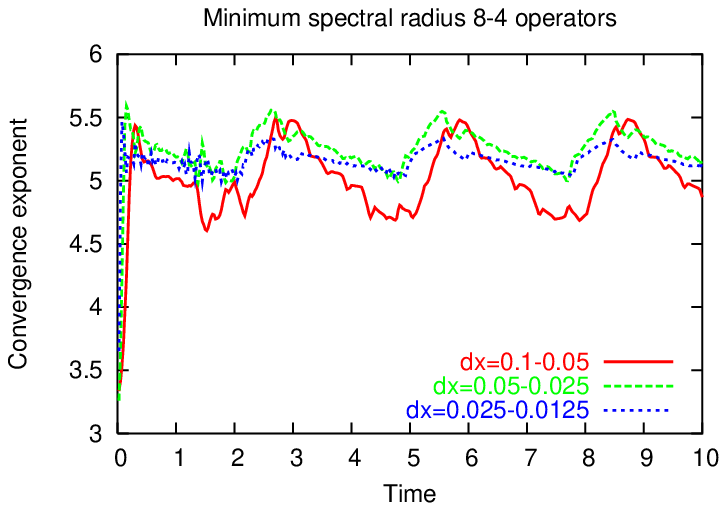}
\includegraphics[width=0.45\textwidth]{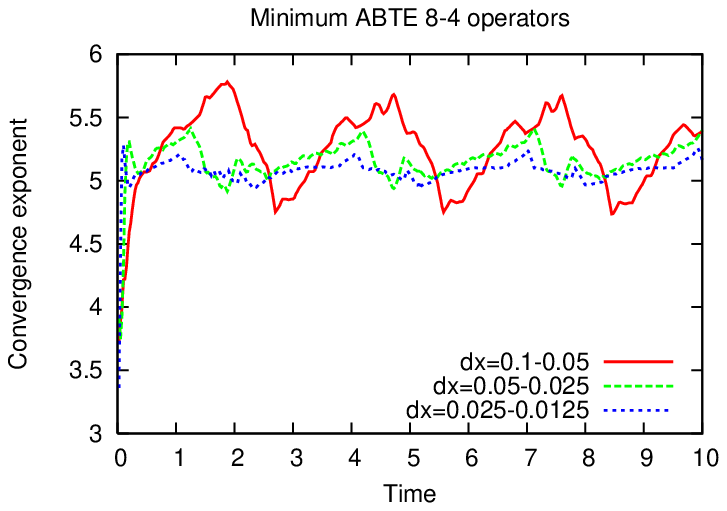}
\caption{Convergence exponents for the minimum spectral radius (top) and the
  minimum ABTE (bottom) $D_{8-4}$ operators.
}
\label{84conv}
\end{center}
\end{figure}

\begin{figure}
\begin{center}
\includegraphics[width=0.45\textwidth]{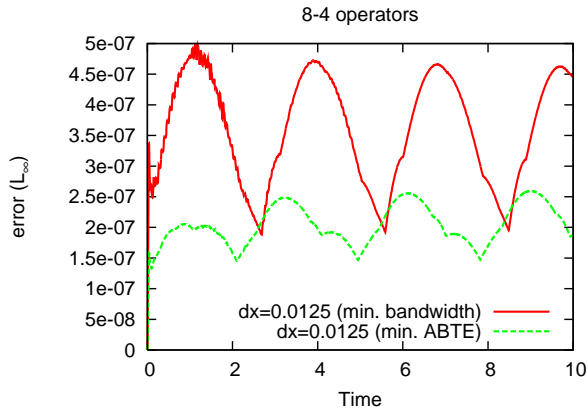}
\caption{Comparison of the accuracy of the two $D_{8-4}$ operator
  types shown in the previous figure in the $L_{\infty}$ norm. Although
  both operators have quite similar error coefficients, there is up to a
  factor of two difference in the errors seen in the actual runs.}
\label{84comp}
\end{center}
\end{figure}

\paragraph{The operator $D_{8-4}$.}
Figure \ref{84conv} shows the results of similar convergence tests, also in
the $L_{\infty}$ norm, for the $D_{8-4}$ operators. 
As discussed in the previous Section, in this case the minimum bandwidth
operator has both very large spectral radius and truncation error coefficients,
so large that it is actually not worthwhile presenting here details of simulations using it 
(they actually crash unless a very
small Courant factor is used, as expected). In references  \cite{Lehner2005a}
and \cite{Svard2005a} 
two different
sets of parameters were
found, both of which reduced the spectral radius by around one order of magnitude, when
compared to the minimum bandwidth one. Here we concentrate on
comparing an operator constructed in a similar way (with slightly smaller
spectral radius than the ones of \cite{Lehner2005a,Svard2005a}) ---that is, minimizing the
spectral radius--- with the minimum ABTE operator. We see that in both cases
we find a global convergence exponent close to five. 
Figure \ref{84comp} shows at fixed resolution (with the highest resolution that we used for the
convergence tests) a comparison between
these two operators, by displaying the errors with respect
to the exact solution, in the $L_{\infty}$ norm. There is an improvement of a
factor of two in the minimum ABTE case (as mentioned, the differences with the minimum bandwidth case
are much larger). 
Notice also that even though not at round-off level, the errors in our simulations are quite small,
of the order of $10^{-7}$ in the $L_{\infty}$ norm (in the $L_2$ norm they are
almost an order of magnitude smaller).

\subsection{Operators based on a restricted full norm}\label{opnondiag}
\subsubsection{Operator properties}

Let us now consider operators that are based on a restricted full norm
(see Section \ref{normdefs}). In this case the norm always depends on the free
parameters, and it is not necessarily positive definite for all values
of them.  Therefore these free parameters are subject to the
constraint of defining a positive definite norm.

We examine here the operators $D_{4-3}$, $D_{6-5}$, and $D_{8-7}$.
These operators have five, seven and nine boundary points,
respectively, and their maximum stencil size are seven, ten and
thirteen points, respectively.

The family of $D_{4-3}$ operators has three independent parameters,
and as mentioned above, they have to be chosen so that the
corresponding norm is positive definite.

We constructed operators by minimizing their bandwidth, their spectral
radius, and their average boundary truncation error. In minimizing the bandwidth
there is some arbitrariness in the choice as to which coefficients in the
stencils are set to zero. Here we follow~\cite{Strand1994a} and set the
coefficients $q_{1,5}$, $q_{2,7}$, and $q_{3,7}$ to zero, where the first index
labels the stencil starting with 1 from the boundary and the second index
labels the point in the stencil. This results in two solutions, but only one
of them corresponds to a positive definite norm.

A comparison of the
spectral radius, ABTE, and error coefficients is listed in table
\ref{tab:4-3-comparison}.
\begin{table}
\caption{Properties of the restricted full norm $D_{4-3}$ operators.}
\begin{ruledtabular}
\begin{tabular}{l|ddd}
& \multicolumn{1}{c}{Minimum} & \multicolumn{1}{c}{Minimum} & \multicolumn{1}{c}{Minimum} \\
Operator & \multicolumn{1}{c}{bandwidth} & \multicolumn{1}{c}{spectral radius} & \multicolumn{1}{c}{ABTE} \\\hline
Spectral radius & 2.428 & 1.322 & 2.758 \\
ABTE & 0.1281 & 0.5824 & 0.0230 \\
$c_1$ & 0.2500 & 1.2359 & -0.0100 \\
$c_2$ & -0.1333 & -0.2378 & -0.0409 \\
$c_3$ & 0.0201 & -0.3166 & 0.0251 \\
$c_4$ & 0.0366 & 0.1053 & -0.0086 \\
$c_5$ & -0.0065 & 0.0279 & -0.0129
\end{tabular}
\label{tab:4-3-comparison}
\end{ruledtabular}
\end{table}

We find in our simulations (see below) that the
minimum bandwidth and the minimum ABTE $D_{4-3}$ operators have very similar
properties.  The latter, however, has slightly smaller errors, enough
to offset the slightly smaller time steps required for stability.  The
operator with a minimum spectral radius unfortunately has very large
errors; in fact, we have not been able to stabilize the system with
any amount of artificial dissipation when we used this operator
for equations with non-constant coefficients in 3D.

The family of $D_{6-5}$ operators has four independent parameters that again
have to be chosen so that the norm is positive definite. For the minimal
bandwidth operator we choose to zero the coefficients $q_{1,7}$, $q_{2,9}$,
$q_{2,10}$ and $q_{3,10}$. The resulting equations cannot be solved
analytically, but numerically we find eight solutions, of which four are
complex. From the remaining real solutions only one of them results in a
positive definite norm. A comparison between
the properties of the minimal bandwidth, minimal spectral radius and minimal
ABTE operators is listed in table~\ref{tab:6-5-comparison}.
\begin{table}
\caption{Properties of the restricted full norm $D_{6-5}$ operators.}
\begin{ruledtabular}
\begin{tabular}{l|ddd}
& \multicolumn{1}{c}{Minimum} & \multicolumn{1}{c}{Minimum} & \multicolumn{1}{c}{Minimum} \\
Operator & \multicolumn{1}{c}{bandwidth} & \multicolumn{1}{c}{spectral radius} & \multicolumn{1}{c}{ABTE} \\\hline
Spectral radius & 2.940 & 1.458 & 3.194 \\
ABTE & 0.0986 & 0.5380 & 0.0648 \\
$c_1$ & 0.1667 & 1.3692 & -0.0154 \\
$c_2$ & -0.1558 & -0.2682 & -0.0507 \\
$c_3$ & 0.0672 & -0.2118 & 0.1336 \\
$c_4$ & 0.0953 & 0.0097 & 0.0532 \\
$c_5$ & -0.0433 & 0.0702 & -0.0733 \\
$c_6$ & 0.0141 & 0.1434 & 0.0187 \\
$c_7$ & -0.0163 & -0.0972 & -0.0123
\end{tabular}
\label{tab:6-5-comparison}
\end{ruledtabular}
\end{table}

As in the $D_{4-3}$ case, the minimum spectral radius operator has a much
smaller spectral radius than the other ones but, again, we did not manage 
to stabilize it with dissipation (at least, with reasonable amounts of it) in the 3D case with non-constant
coefficients. The errors for the minimum ABTE operator are significantly smaller
than the minimum bandwidth one, which is reflected in practice by
smaller errors; in addition, the operator could be stabilized with significantly less
artificial dissipation.

It should be noted at this point that we did not manage to stabilize
the $D_{6-5}$ operator with a naive adapted Kreiss--Oliger like
dissipation prescription.  We tried applying Kreiss--Oliger
dissipation in the interior of the domain, and applying no dissipation
near the boundary where the centered stencils cannot be applied.  This
failed. Presumably this is caused by the fact that the dissipation operator is not negative
semi-definite with respect to the SBP norm. Only after constructing boundary
dissipation operators following the approach of Ref. \cite{Mattsson2004a}, did we arrive at a stable
scheme involving the $D_{6-5}$ operator.

The family of $D_{8-7}$ operators has five independent parameters. When
attempting to obtain minimum bandwidth operators by setting the coefficients
$q_{1,9}$, $q_{2,11}$, $q_{2,12}$, $q_{2,13}$, and $q_{3,13}$ to zero,
we numerically
find 24 solutions, of which 16 are complex and 8 are real. However, none of
these solutions yields a positive definite norm. The minimum spectral radius
operator has so large error coefficients that we could not stabilize it with
reasonable amounts of dissipation in  
the 3D non-constant coefficient case. The minimum ABTE operator has a spectral
radius larger than 60000 and so would require very small time steps when
explicit time evolution is used.

Since none of the operators considered
so far was usable in practice, we experimented with several other ways of
choosing the parameters. First, we tried to minimize a weighted
average of the spectral radius and ABTE for different weight values,
but none of these operators turned out to be useful, even though
their properties were much improved. By minimizing the sum of the squares
of the difference between error coefficients in neighboring boundary points,
we next attempted to reduce the noise produced in the boundary region. Even
when weighted with the spectral radius, these operators proved not to be an
improvement. Finally, based on the observation that some choices of parameters
lead to very large values in the inverse of the norm, which directly affects the
dissipation operator near the boundary (see Eq.\ (\ref{dismat})),
we speculated that for some parameter
sets the dissipation operator might make things worse near the boundary, and
experimented with choosing parameters that would minimize the condition number
of the norm in combination with any of the previously mentioned properties.

However, even though we were able to find parameter sets that
looked reasonable with respect to spectral radius, error coefficients and
properties of the norm, none of the operators that we constructed could
be stabilized with any amount of dissipation in the 3D non-constant coefficient
case. This is presumably related to some important properties not holding in
the non-diagonal case, as discussed in Section \ref{diag_vs_nondiag}. 
Of course, it could still happen that usable $D_{8-7}$ operators do
exist, using some other criteria to choose parameters. 

\subsubsection{Numerical tests}  \label{numnondiag}

For the restricted full operators we usually have to add
dissipation.  There are several causes for this, which are well
understood; we have reviewed them in Section \ref{setup}.

\paragraph{The operator $D_{4-3}$.}
One main driving point behind using operators based on non-diagonal
norms is that their order of accuracy near the boundary is higher.
Our operators based on restricted full norms drop one order of
accuracy near the boundary, which means that the global convergence
order is, in theory, not affected.  (See the standard $D_{6-3}$
operator, described in Section \ref{D_6_3}, and illustrated in Figure
\ref{63conv}, for an example where this is false in practice.)

\begin{figure}
\begin{center}
\includegraphics[width=0.45\textwidth]{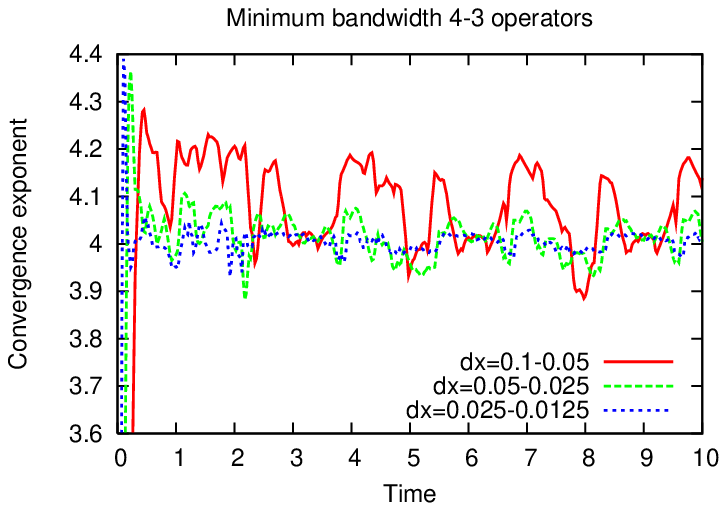}
\includegraphics[width=0.45\textwidth]{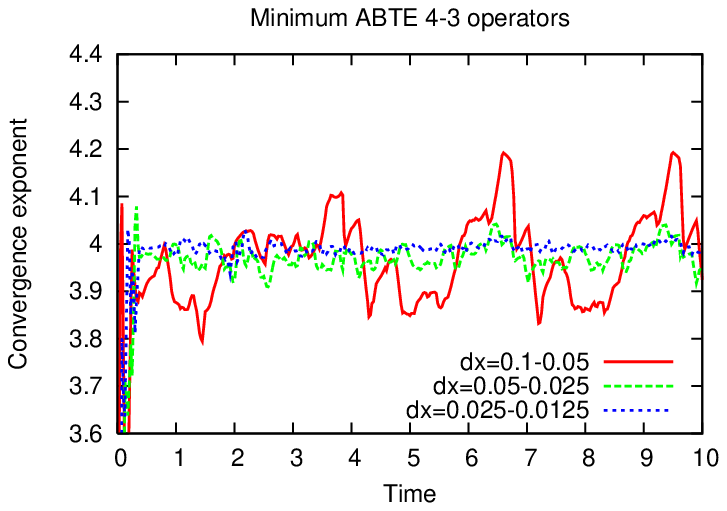}
\caption{Convergence exponents for the minimum bandwidth (top) and the
  minimum ABTE (bottom) $D_{4-3}$ operators.}
\label{43conv}
\end{center}
\end{figure}

Figure \ref{43conv} shows the results of convergence tests in the
$L_{\infty}$ norm for $D_{4-3}$ operators, for both the minimum
bandwidth and the new, optimized, minimum
ABTE operator. In both cases the global
convergence exponent is very close to four.  Different from the
operators based on diagonal norms, the convergence order is also
almost constant in time.  This may be so because the boundary is here
not the main cause of discretization error, so that the resulting
accuracy is here independent of what kind of feature of the solution
is currently propagating through the boundary.

\begin{figure}
\begin{center}
\includegraphics[width=0.45\textwidth]{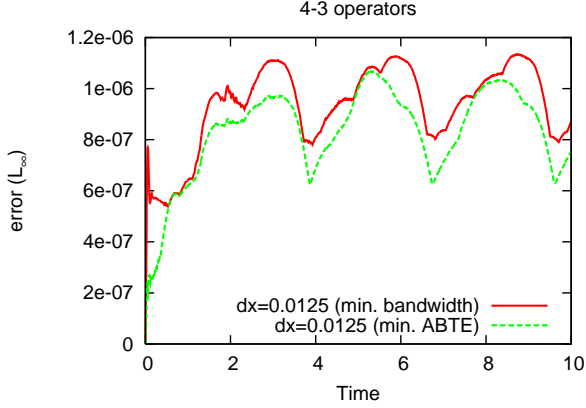}
\caption{Accuracy comparison of the two types of $D_{4-3}$ operators
  shown in the previous figure in the $L_{\infty}$ norm. Contrary to the
  results for the $D_{6-3}$ and $D_{8_4}$, here the difference in errors
  is minimal.}
\label{43comp}
\end{center}
\end{figure}

\begin{figure}
\begin{center}
\includegraphics[width=0.45\textwidth]{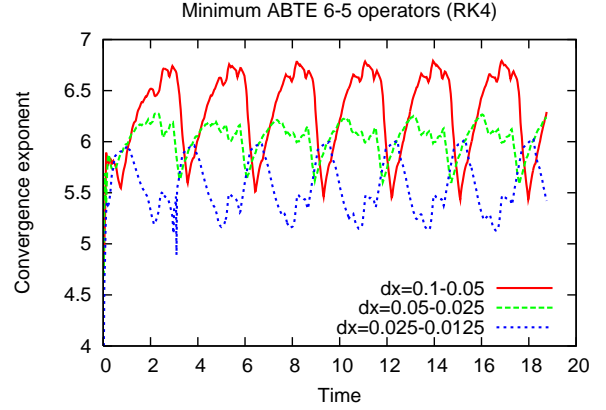}
\includegraphics[width=0.45\textwidth]{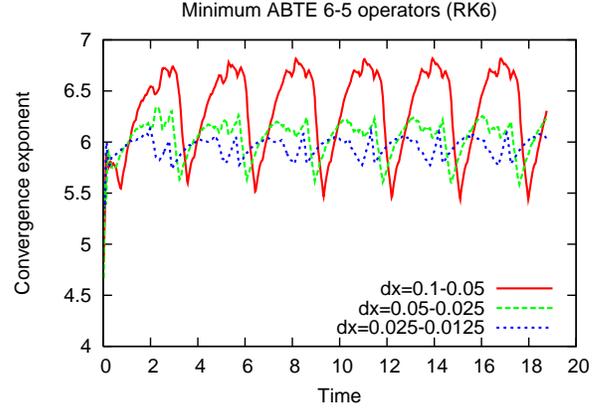}
\caption{Convergence for the optimized $D_{6-5}$ operator with dissipation and
  Runge--Kutta time integrators of order four (top) and six (bottom),
  respectively.  We see a slightly lower convergence order for the
  highest resolution when RK4 is used.  This effect is not present
  with the RK6 integrator. The lower convergence order indicates that
  the accuracy of the spatial finite differencing operators is high
  enough, so that the overall error is dominated by the accuracy of the
  time integrator.}
\label{65conv}
\end{center}
\end{figure}

Figure \ref{43comp} compares the two $D_{4-3}$ operators at a fixed,
high resolution.  The two operators lead to very similar $L_{\infty}$
norms of the solution error.  This difference in accuracy is much
smaller than it was for the operators based on diagonal norms.
Altogether, the optimization leads to no practical advantage.
\paragraph{The operator $D_{6-5}$.}
The operator $D_{6-5}$ is expected to have the highest global order of
accuracy of all the operators discussed in this paper (since we were
not able to stabilize the $D_{8-7}$ operator with dissipation). We do in fact
see sixth order convergence, but only when using a sufficiently
accurate time integration scheme. 
Figure \ref{65conv} shows the results of convergence tests in the
$L_{\infty}$ norm for the new, optimized $D_{6-5}$ operator with
dissipation, using a
Courant factor $\lambda = 0.25$, for both fourth and sixth order
accurate Runge--Kutta time integrators (RK4 and RK6).

When RK4 is used, the convergence exponent drops from about $6$ to
close to $5$ for our highest resolution.  This effect is not present
when we use RK6, indicating that it is the time integrator's lower
convergence order that actually poisons the results when resolution is
increased.
Instead of using a higher order time integrator, it would also have
been possible to reduce the time step size.  Especially in complicated
geometries, an adaptive step size control is very convenient; this
lets one specify the desired time integration error, and the step size
is automatically adjusted accordingly.

\subsection{Comparison between operators}\label{comp}
\subsubsection{Comparison of accuracy}
\begin{figure}
\begin{center}
\includegraphics[width=0.45\textwidth]{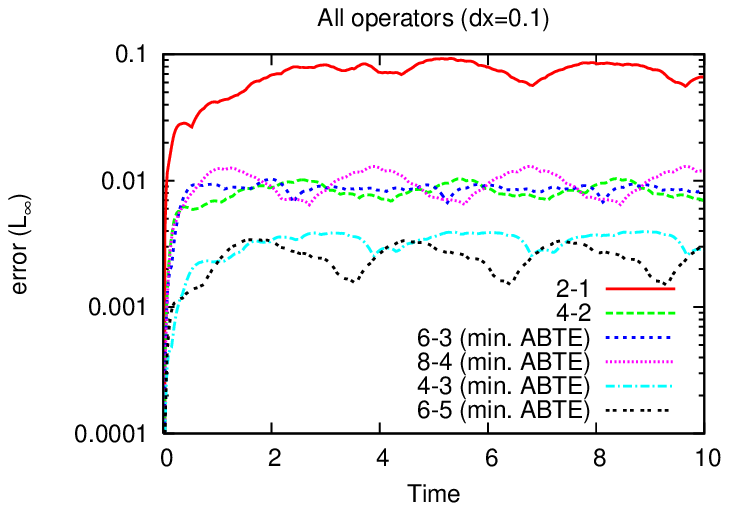}
\includegraphics[width=0.45\textwidth]{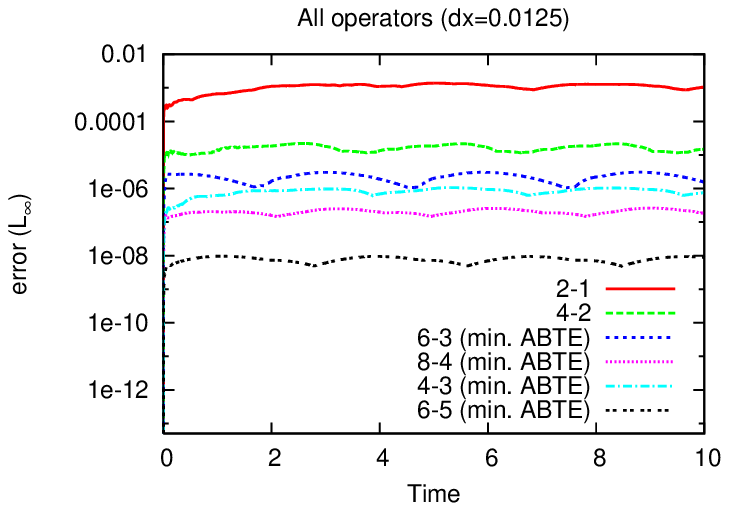}
\includegraphics[width=0.45\textwidth]{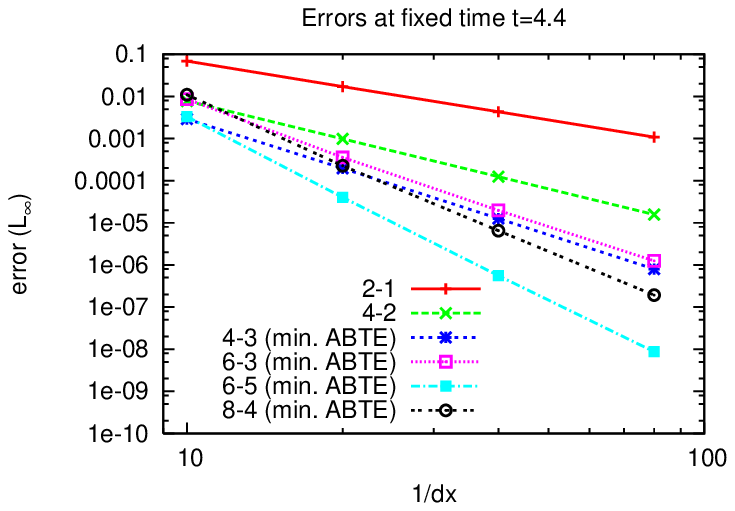}
\caption{Comparison of the errors for all the new, optimized, differencing operators
  constructed in this paper. The top (middle) plot shows a comparison of
  all the unique and optimized operators at low (high) resolutions. The
  bottom plot shows the error at $t=4.4$ for all resolutions. The
  most successful operators are the optimized $D_{6-5}$, $D_{4-3}$,
  and $D_{8-4}$.}
\label{comparison}
\end{center}
\end{figure}

We now compare the operators based on a diagonal and on a restricted
full norm that we described and examined above.  Figure
\ref{comparison} shows, for two different resolutions, the solution
errors for all our new operators.  As one can see, our best performing
operators are $D_{8-4}$ and $D_{6-5}$ . One can also see that, for the
highest resolution shown there, which corresponds to $161^3$ grid
points per block, there is a difference of \emph{five orders of
  magnitude} between the errors of $D_{6-5}$ and $D_{2-1}$.  This
demonstrates nicely the superiority of our new high order operators
when a high accuracy is desired.

Even for the lowest resolution shown here, which uses only $21^3$ grid
points per block, the difference is still more than one order of
magnitude, indicating that $D_{6-5}$ or $D_{4-3}$ would be fine
choices there.

\subsubsection{Comparison of cost}

Higher order operators allow higher accuracy with the same number of
grid points, but they also require more operations per grid point and
thus have a higher cost.  Table \ref{tab:costs} compares actual run
times per time integrator step for runs with $161^3$ grid points per block.
These measured costs include not only calculating the right hand side,
which requires taking the derivatives, but also applying the boundary
conditions, which requires decomposing the system into its
characteristic modes, and also some inter-processor communication.
The time spent in the time integration itself is negligible.
\begin{table}
  \caption{
    Relative operator costs for our test problem, measured in seconds
    per time step with an RK4 integrator.  The number in parentheses
    for the $D_{6-5}$ used an RK6 time integrator, which requires seven
    iterations instead of four.  The
    numbers marked with an asterisk were obtained from runs using a
    larger number of processors and are probably larger than they
    would be otherwise due to scaling inefficiencies.
  }
  \begin{ruledtabular}
    \begin{tabular}{l|dd}
      {}       & \multicolumn{1}{c}{Time [sec]}
      & \multicolumn{1}{c}{Time [sec]}  \\
      Operator & \multicolumn{1}{c}{without dissipation}
      & \multicolumn{1}{c}{with dissipation} \\\hline
      $D_{2-1}$ & 26.9 & 27.7 \\
      $D_{4-2}$ & 27.9 & 29.7 \\
      $D_{6-3}$ & 30.5 & 32.0 \\
      $D_{8-4}$ & 34.4 & 35.3 \\
      $D_{4-3}$ & 28.6 & 44.8 \\
      $D_{6-5}$ & 41.6^* & 39.1\; (86.2^*)
    \end{tabular}
    \label{tab:costs}
  \end{ruledtabular}
\end{table}

Larger stencils increase the cost slightly.  This increase in cost is
more than made up by the increase in accuracy, as shown above.  For
operators based on diagonal norms, adding dissipation to the system
increases the cost only marginally.  The dissipation operators for
derivatives based on non-diagonal norms are more complex to calculate
and increase the run time noticeably.  All numbers were obtained
from a straightforward legible implementation in Fortran, without
spending much effort on optimizing the code for performance.

The effect of higher order operators on the overall run time is not
very pronounced.  As they do not increase the storage
requirements either, the only reason speaking against using them (for smooth
problems) seems to be the effort one has to spend constructing and
implementing them.

\section{Conclusion} \label{end}

Let us summarize the main points of this paper.  We have explicitly
constructed accurate, high-order finite
differencing operators which satisfy summation by parts.  This
construction is not unique, and it is necessary to specify some free
parameters.  We have considered several optimization criteria to define
these parameters; namely, (a) a minimum bandwidth, (b) a minimum of the
truncation error on the boundary points, (c) a minimum spectral
radius, and a combination of these. 

We examined in detail a set of operators that are up to
tenth order accurate.  We found that minimum bandwidth operators may
have a large spectral radius or truncation error near the boundary.
Optimizing for these two criteria can, surprisingly, reduce the
operators' spectral radius by three orders
and their accuracy near the boundary
by two orders of magnitude.  

Some of the finite differencing operators require artificial
dissipation.  We have therefore also constructed high-order
dissipation operators, compatible with the above finite differencing
operators, and also semi-definite with respect to the SBP scalar product.

We tested the stability and accuracy of these operators by evolving a
scalar wave equation on a spherical domain. Our domain is split into
seven blocks, each discretized with a structured grid.  These blocks
are connected through penalty boundary conditions.  We demonstrated
that the optimized finite differencing operators have also far
superior properties in practice.  The most accurate operators are
$D_{8-4}$ and $D_{6-5}$; the latter is in our setup five orders of
magnitude more accurate than a simple $D_{2-1}$ operator.

\begin{acknowledgments}
We are deeply indebted to Jos\'{e} M. Mart\'{\i}n-Garc\'{\i}a for allowing
us to use his well organized Mathematica notebook to construct the
SBP operators, 
and to Mark Carpenter, Heinz-Otto Kreiss, Ken Mattsson, and Magnus
Sv\"ard for numerous helpful discussions and suggestions.
We also thank Luis Lehner, Harald Pfeiffer, Jorge Pullin,
and Olivier Sarbach for discussions,
suggestions and comments on
the manuscript, and Cornell University and the Albert Einstein
Institute for hospitality at different stages of this work.
As always, our numerical calculations would have been impossible
without the large number of people who made their work available to
the public: we used the Cactus computational toolkit
\cite{Goodale02a,cactusweb1} with a number of locally developed
thorns, the LAPACK \cite{lapackweb} and BLAS \cite{blasweb} libraries
from the Netlib Repository \cite{netlibweb}, and the LAM
\cite{burns94:_lam, squyres03:_compon_archit_lam_mpi, lamweb} and
MPICH \cite{Gropp:1996:HPI, mpich-user, mpichweb} MPI \cite{mpiweb}
implementations.
 E. Schnetter acknowledges funding from the DFG's
special research centre TR-7 ``Gravitational Wave Astronomy''
\cite{sfbtr7web}. 
This research was supported in part by the NSF under Grant PHY0505761 and NASA under Grant NASA-NAG5-1430
to Louisiana State University, by the NSF under Grants PHY0354631 and PHY0312072 to
Cornell University, by the
National Center for Supercomputer Applications under grant MCA02N014 and
utilized Cobalt and Tungsten, it used resources of the National Energy
Research Scientific Computing Center, which is supported by the Office of
Science of the U.S. Department of Energy under Contract No.\ DE-AC03-76SF00098
and it employed the resources of the Center
for Computation and Technology at Louisiana State University, which is
supported by funding from the Louisiana legislature's Information
Technology Initiative. 
\end{acknowledgments}

\appendix
\section{Operator coefficients}
We provide, for the reader's convenience, the coefficients for the
derivative and dissipation operators that we constructed above.  Since
the values of these coefficients themselves are only of limited
interest, and since there is a large danger of introducing errors when
typesetting these coefficients, we make them available electronically
instead.  We also make a Cactus \cite{cactusweb1} thorn
\emph{SummationByParts} available via anonymous CVS.  This thorn
implements the derivative and dissipation operators.  Our web pages
\cite{numrelcctweb} contain instructions for accessing these.  For the
sake of continuity, we will also make the coefficients available on www.arxiv.org
together with this article.

We distribute the coefficients as a set of files, where each file
defines on operator.  The content of the file is written in a
Fortran-like pseudo language that defines the coefficients in
declarations like
\begin{verbatim}
  a(1) = 0.5
  q(2,3) = 42.0
\end{verbatim}
and sometimes makes use of additional constants, as in
\begin{verbatim}
  x1 = 3
  a(2) = x1 + 4
\end{verbatim}
We write here \verb+a(1)+ as $a_1$ and \verb+q(2,3)+ as $q_{23}$.

\subsection{Derivative operators}
The derivative operators $D_{2-1}$, $D_{4-2}$, $D_{6-3}$, $D_{8-4}$,
$D_{4-3}$, and $D_{6-5}$ are defined via coefficients $a_i$ and
$q_{ij}$.  In the interior of the domain it is
\begin{eqnarray}
  D_{ij} u_j & = & \frac{1}{h} \sum_{j=1}^s a_j \left( u_{i+j} -
    u_{i-j} \right)
\end{eqnarray}
and near the left boundary it is (i.e.\ $i=1,b$)
\begin{eqnarray}
  D_{ij} u_j & = & \frac{1}{h} \sum_{j=1}^s q_{ji} u_j.
\end{eqnarray}
At the right boundary the same coefficients are used in opposite order and
with opposite sign.
  
\subsection{Dissipation operators} \label{dis_ap}
\subsubsection{Dissipation operators based on diagonal norms}
The dissipation operators corresponding to $D_{2-1}$, $D_{4-2}$,
$D_{6-3}$, and $D_{8-4}$ are defined via coefficients $a_{ij}$ and
$q_{i}$.%

In the interior of the domain it is
\begin{eqnarray}
  A_{ij} u_j & = & \frac{\epsilon}{2^{2p}} \left[ q_0 u_i +
    \sum_{j=1}^s q_{j} \left( u_{i-j} + u_{i+j} \right) \right]
\end{eqnarray}
and near the boundary it is
\begin{eqnarray}
  A_{ij} u_j & = & \frac{\epsilon}{2^{2p}} \sum_{j=1}^s a_{ji} u_j, 
\end{eqnarray}
where $\epsilon\ge 0$ selects the amount of dissipation and is usually of
order unity.

\subsubsection{Dissipation operators based on non-diagonal norms}
The dissipation operators corresponding to $D_{4-3}$ and
$D_{6-5}$\ldots are more complicated, since they depend on the user
parameters specifying the number of grid points, $N$, and the size of the
transition region (i.e.\ the region where $B_p$ is different from 1).

The dissipation operators are then constructed according to Eq.~(\ref{dismat})
\begin{equation}
  A_{2p} = - \frac{\epsilon}{2^{2p}}\, h^{2p}\; \Sigma^{-1} D_p^T B_p D_p.
\end{equation}

The coefficients for the inverse of the norm, $\Sigma^{-1}$, are provided
in the boundary region only.  In the files this inverse is denoted by
\verb+sigma(i,j)+.  In the interior the norm (and its inverse) is
diagonal with value $1$.

In the $D_{4-3}$ case, the $N\times N$ matrix $D_2$ defining the consistent
approximation of $d^2/dx^2$ is given by
\begin{equation}
D_{2} = \frac{1}{h^2} \left ( \begin{array}{rrrrcrrrr}
  1 & -2 & 1  & 0 &        &   &    &    &   \\
  1 & -2 & 1  & 0 &        &   &    &    &   \\
  0 & 1  & -2 & 1 &        &   &    &    &   \\
    &    &    &   & \ddots &   &    &    &   \\
    &    &    &   &        & 1 & -2 & 1  & 0 \\
    &    &    &   &        & 0 & 1  & -2 & 1 \\
    &    &    &   &        & 0 & 1  & -2 & 1
  \end{array} \right ),
\end{equation}
while the diagonal matrix $B_2$ has the value $h$ at either boundary and
increases linearly to the value 1 across the user defined transition region.

In the $D_{6-5}$ case the matrix defining the consistent
approximation of $d^3/dx^3$ near the left boundary $D_3^l$ is
\begin{equation}
D_3^l = \frac{1}{h^3} \left ( \begin{array}{rrrrrc}
  -1 &  3 & -3 &  1 & 0 &       \\
  -1 &  3 & -3 &  1 & 0 &       \\
  -1 &  3 & -3 &  1 & 0 &       \\
   0 & -1 &  3 & -3 & 1 &        \\
     &    &    &    &   & \ddots
  \end{array} \right ),
\end{equation}
while at the right boundary $D_3^r$ is
\begin{equation}
D_3^r = \frac{1}{h^3} \left ( \begin{array}{crrrrr}
  \ddots &    &    &    &    &   \\
         & -1 &  3 & -3 &  1 & 0 \\
         &  0 & -1 &  3 & -3 & 1 \\
         &  0 & -1 &  3 & -3 & 1 \\
         &  0 & -1 &  3 & -3 & 1
  \end{array} \right ).
\end{equation}
Since the values on the diagonal in the interior of $D_3^l$ and $D_3^r$ are
$-3$ and $3$, respectively, it is impossible to construct
a single matrix $D_3$ to cover the whole domain. However, since both matrix
products $(D_3^l)^T D_3^l$ and $(D_3^r)^T D_3^r$ result in the same interior
operator, dissipation operators can be constructed in the left and right
domain separately and then combined into a global operator.
The diagonal matrix $B_3$ has the values $h^2$ at the boundary and 
and $1$ in the interior, and a third order polynomial with zero
derivative at either end of the transition region is used to smoothly connect
the boundary with the interior.
\bibliographystyle{apsrev}
\bibliography{bibtex/references}
\end{document}